\documentclass[12pt]{article}
\begin{document}
\title{Noncommutative Spacetime Effects and Gravitation}
\author{B.G. Sidharth\\
International Institute for Applicable Mathematics \& Information Sciences\\
Hyderabad (India) \& Udine (Italy)\\
B.M. Birla Science Centre, Adarsh Nagar, Hyderabad - 500 063 (India)}
\date{}
\maketitle
\begin{abstract}
In recent years Quantum Superstrings and Quantum Gravity approaches have come to rely on non differenciable spacetime manifolds. These throw up a noncommutative spacetime geometry and we consider the origin of mass and a related modification of the Dirac equation in this context. This also throws some light on gravitation itself.
\end{abstract}
\section{Origin of Mass}
In an earlier communication \cite{ijmpe} it was shown that the effect of noncommutative geometry in breaking the symmetry in non-Abelian gauge theory leads to a term identical to the Higgs boson generating term. Indeed over the past few years various Quantum Gravity schemes as also string theoretic approaches are converging to the fact that spacetime is not smooth, but rather there is a minimum cut off which leads to a noncommutative geometry \cite{hooft,hooft2,bgs,bgs2,univfl,witten,neeman}. On the other hand it has been argued by the author that inertial mass is a result of non local Quantum amplitudes at the Compton scale \cite{ijpap,cu}. We will now reconcile both these concepts and show that the mass term appears as a result of the noncommutativity of spacetime. Our starting point is the Dirac equation (using natural units $c = 1 = \hbar$),
\begin{equation}
\left(\gamma^\mu p_\mu\right) \psi = 0\label{e1}
\end{equation}
Remembering that the operator in (\ref{e1}) is
$$\gamma^\circ p^\circ - \vec{\gamma}_\circ \vec{p},$$
we multiply on the left side by
$$\gamma^\circ p^\circ + \vec{\gamma}_\circ \vec{p}$$
This gives us
\begin{equation}
\left[\left(p^2_0 - \vec{p}^2\right) - \imath \vec{\sum}\cdot \left(\vec{p}\times \vec{p}\right) + \gamma^\imath \gamma^\circ B_{\imath \circ}\right]\psi = 0\label{e2}
\end{equation}
where $\vec{\sum} = \left(\begin{array}{ll}
\vec{\sigma} \quad 0 \\
0 \quad \vec{\sigma}
\end{array}
\right)$. In (\ref{e2}), the first term is the usual energy momentum term which leads to the massless Klein-Gordon or D'Alembertian operator. The second term is well known (Cf.ref.\cite{schiff}) - in a non relativistic approximation with a small external magnetic field that is switched on, this term leads to a spin orbit coupling. It is the third and last term that is the extra effect due to the noncommutative character of spacetime, that is due to the fact
\begin{equation}
B_{\mu \nu} = \left[p_\mu , p_\nu \right] \ne 0\label{e3}
\end{equation}
We identify this extra term with the mass term, viz.,
\begin{equation}
\gamma^\imath \gamma_\circ B_{\imath \circ} = m^2\label{e4}
\end{equation}
We will justify this identification in a moment. With this identification, and in the absence of an external magnetic field, in which case the second term in (\ref{e2}) disappears, (\ref{e2}) goes over to the Klein-Gordon equation for a massive particle.\\
Infact it has already been discussed in detail (Cf.ref.\cite{bgs2,nc116,springer} that in the above noncommutative case, (\ref{e3}),
\begin{equation}
B_{\mu \nu} = p_\mu p_\nu - p_\nu p_\mu = \partial_\mu \partial_\nu - \partial_\nu \partial_\mu = e F_{\mu \nu}\label{e5}
\end{equation}
where $F_{\mu \nu}$ is the usual electromagnetic field tensor. (The deep relation of (\ref{e5}) with the Weyl gauge invariant electromagnetic potential has also been discussed in detail in the above references). Because of (\ref{e5}) and because of the fact that,
$$
\gamma^\imath \gamma^\circ = \alpha^\imath$$
where $\vec{\alpha}$ denotes the velocity operator, at the Compton wavelength where momentum is $m (= mc)$, the extra term becomes
$$\frac{e^2}{l^2} \sim m^2$$
in agreement with (\ref{e4}) due to the definition of the Compton length as the electron radius viz.,
$$l \sim e^2/m \sim \frac{1}{m}$$
The massive Klein-Gordon equation then, in the usual formulation leads back to the Dirac equation (\ref{e1}) but this time with the usual mass term.
\section{An Ultra High Energy Dirac Equation}
As noted over the past few years, different approaches towards Quantum Gravity are leading to the conclusion that spacetime is not a smooth continuum but rather, has a discrete structure. Indeed as 't Hooft put it \cite{hooft,univfl}, ``It is somewhat puzzling to the present author why the lattice structure of space and time had escaped attention from other investigators up till now...''. Such a discrete structure would imply a violation of Lorentz symmetry at ultra high energies, as noted by the author himself, Glashow, Coleman and several others \cite{ijtp}.\\
Indeed some of this work has been motivated by the possibility that such a violation has already been observed in a few cases in a study of ultra high energy cosmic rays (Cf.ref.\cite{ijtp} and several references therein). This has the consequence that the Klein-Gordon equation gets modified as noted in \cite{ijtp}. We will now consider the modification in the Dirac equation and briefly examine its consequences.\\
Once we consider a discrete spacetime structure, the energy momentum relation, as noted, gets modified \cite{cu,mont} and we have in units $c = 1 = \hbar$,
\begin{equation}
E^2 - p^2 - m^2 + l^2 p^4 = 0\label{e1a}
\end{equation}
$l$ being a minimum length interval, which could be the Planck length or more generally the Compton length. Let us now consider the Dirac equation
\begin{equation}
\left\{ \gamma^\mu p_\mu - m\right\} \psi \equiv \left\{\gamma^\circ p^\circ + \Gamma \right\} \psi = 0\label{e2a}
\end{equation}
If we include the extra effect shown in (\ref{e1a}) we get
\begin{equation}
\left(\gamma^\circ p^\circ + \Gamma + \beta l p^2\right) \psi = 0\label{e3a}
\end{equation}
$\beta$ being a suitable matrix.\\
Multiplying (\ref{e3a}) by the operator 
$$\left(\gamma^\circ p^\circ - \Gamma - \beta l p^2\right)$$
on the left we get
\begin{equation}
p^2_0 - \left(\Gamma \Gamma + \left\{\Gamma \beta + \beta \Gamma\right\} + \beta^2 l^2 p^4\right\} \psi = 0\label{e4a}
\end{equation}
If (\ref{e4a}), as in the usual theory, has to represent (\ref{e1a}), then we require that the matrix $\beta$ satisfy 
\begin{equation}
\Gamma \beta + \beta \Gamma = 0, \quad \beta^2 = 1\label{e5a}
\end{equation}
From the properties of the Dirac matrices \cite{bd} it follows that (\ref{e5a}) is satisfied if
\begin{equation}
\beta = \gamma^5\label{e6a}
\end{equation}
Using (\ref{e6a}) in (\ref{e3a}), the modified Dirac equation finally becomes
\begin{equation}
\left\{\gamma^\circ p^\circ + \Gamma + \gamma^5 l p^2\right\} \psi = 0\label{e7a}
\end{equation}
Owing to the fact that we have \cite{bd}
\begin{equation}
P \gamma^5 = -\gamma^5 P\label{e8a}
\end{equation}
It follows that the modified Dirac equation (\ref{e7a}) is not invariant under reflections. This is a result which is to be expected because the correction to the usual energy momentum relation, as shown in (\ref{e1a}) arises when $l$ is of the order of the Compton wavelength. The usual Dirac four spinor 
$\left(\begin{array}{ll}\Theta \\ \chi
\end{array}
\right)$ as is known has the so called positive energy (or large) components $\Theta$ and the negative energy (or small) components $\chi$. However as is well known, when we approach the Compton wavelength, that is as 
$$p \to mc$$
the roles are reversed and it is the $\chi$ components which predominate. Moreover the $\chi$ two spinor behaves under reflection as \cite{bd}
$$\chi \to -\chi$$
We can also see that due to the modified Dirac equation (\ref{e7}), there is no additional effect on the anomalous gyromagnetic ratio. This is because, in the usual equation from which the magnetic moment is determined \cite{merz} viz.,
$$\frac{d\vec{S}}{dt} = -\frac{e}{\mu c} \vec{B} \times \vec{S},$$
where $\vec{S} = \hbar \sum/2$ is the electron spin operator, there is now an extra term
\begin{equation}
\left[\gamma^5, \sum\right]\label{e9a}
\end{equation}
However the expression (\ref{e9a}) vanishes by the property of the Dirac matrices.\\
It has already been argued in detail that \cite{ijpap,cu} as we approach the Compton wavelength, the Dirac equation describes the quark with the fractional charge and handedness. Our above derivation and conclusioin is pleasingly in agreement with this result.
\section{Gravitation}
We now come to a slightly different problem-- that of Gravitation.
Gravitation has defied a unification with electromagnetism for nearly nine decades now. In the context of Quantum Field Theory, the fact that the graviton is a spin 2 particle makes it difficult for a description in terms of gauge field theory. As Witten put it \cite{witten} ``The existence of gravity clashes with our description of the rest of physics by quantum fields.'' On the other hand Wolfgang Pauli went so far as to mention that we must not try to put together what God had intended to be separate. However, it has been shown that in the context of sections 1 and 2, it is still possible to effect a unified description of Gravitation and Electromagnetism \cite{bgs,univfl,cu,nc116}.  In a slightly different context, we consider below a simplified model which shows up gravitation as a residual electrical effect.\\
Our starting point is the  well known fact that the universe at large is electrically neutral, in the sense that it is dominated by atoms in which the positive and negative electric charges neutralise each other. This is the reason why electric charges have a marginal role in the large scale universe, the much weaker gravitation being the predominant force.\\
With this background let us consider the following simple model of an electrically neutral atom which nevertheless has a dipole effect. Infact as is well known from elementary electrostatics the potential energy at a distance $r$ due to the dipole is given by
\begin{equation}
\phi = \frac{\mu}{r^2}\label{ea1}
\end{equation}
where $\mu = eL, L \sim 10^{-8}cm \sim 10^3 l = \beta l,$ $e$ being the electric charge of the electron for simplicity and $l$ being the electron Compton wavelength. (There is a factor $cos \Theta$ with $\mu$, but on an integration over all directions, this becomes an irrelevant constant factor $4 \pi$.)\\
Due to (\ref{ea1}), the potential energy of a proton $p$ (which approximates an atom in terms of mass) at the distance $r$ (much greater than $L$) is given by
\begin{equation}
\frac{e^2L}{r^2}\label{ea2}
\end{equation}
As there are $N \sim 10^{80}$ atoms in the universe, the net potential energy of a proton due to all the dipoles is given by
\begin{equation}
\frac{Ne^2L}{r^2}\label{ea3}
\end{equation}
In (\ref{ea3}) we use the fact that the predominant effect comes from the distant atoms which are at a distance $\sim r$, the radius of the universe.\\
We now use the well known Eddington formula,
\begin{equation}
r \sim \sqrt{N} l\label{ea4}
\end{equation}
$r$ being of the order of the dimension of the universe. (Incidentally while (\ref{ea4}) has been known as an empirical relation for nearly a century now, it can actually be deduced from theory \cite{ijmpa,psp,cu}). If we introduce (\ref{ea4}) in (\ref{ea3}) we get, as the energy $E$ of the proton under consideration
\begin{equation}
E = \frac{\sqrt{N}e^2\beta}{r}\label{ea5}
\end{equation}
Let us now consider the gravitational potential energy $E'$ of the proton $p$ due to all the other $N$ atoms in the universe. This is given by
\begin{equation}
E' = \frac{GMm}{r}\label{ea6}
\end{equation}
where $m$ is the proton mass and $M$ is the mass of the universe.\\
Comparing (\ref{ea5}) and (\ref{ea6}), not only is $E$ equal to $E'$, but remembering that $M = Nm$, we get
\begin{equation}
\frac{e^2}{Gm^2} = \frac{1}{\sqrt{N}}\label{ea7}
\end{equation}
Equation (\ref{ea7}) has been a well known relatioin for many decades \cite{mwt}. It gives the ratio of the electromagnetic and gravitational coupling constants.\\
We can now argue that the significance of (\ref{ea7}) and the equality of (\ref{ea5}) and (\ref{ea6}) is the following: The gravitational energy itself is a manifestation of the residual dipole effect of the electric charges which constitute the atoms of the universe. Indeed, it has been shown \cite{ijmpa,cu} that
\begin{equation}
G = \frac{lc^2}{m\sqrt{N}}\label{ea8}
\end{equation}
And it was argued \cite{fpl,univfl} that (\ref{ea8}) is symptomatic of the fact that gravitation, unlike electromagnetism is a distributional effect, distributed over the $N$ particles of the universe. Our conclusion in (\ref{ea7}) is pleasingly in agreement with this conclusion.

\end{document}